\documentclass[%
 aip,
 amsmath,amssymb,
 reprint,%
]{revtex4-1}

\usepackage{graphicx}% Include figure files
\usepackage{subfigure}
\usepackage{dcolumn}% Align table columns on decimal point
\usepackage{bm}% bold math
\usepackage[utf8]{inputenc}
\usepackage[T1]{fontenc}
\usepackage{mathptmx}
\usepackage{amssymb}
\usepackage{epstopdf}

\usepackage[breaklinks=true,
colorlinks=true,
linkcolor=blue,
citecolor=blue,
urlcolor=blue]{hyperref}

\begin{document}

\preprint{AIP/123-QED}

\title{A Wave-Based Simulation Model for Cross-Beam Energy Transfer and Stimulated Brillouin Scattering in Laser-Plasma Systems}
% Force line breaks with \\

\author{Y. Chen}
\affiliation{School of Electrical and Information Engineering, Anhui University of Science and Technology, Huainan, Anhui 232001, China}
\author{Qing Wang}
\affiliation{Institute of Applied Physics and Computational Mathematics, Beijing, 100094, China}
\author{H. Wen } %{$^{\dagger}$}
  \affiliation{ School of Physics and Electronics, Hunan University, Changsha, 410082, China}
\author{Y. Z. J. Xu}
\affiliation{School of Electrical and Information Engineering, Anhui University of Science and Technology, Huainan, Anhui 232001, China}
\author{S. J. Peng}
\affiliation{School of Electrical and Information Engineering, Anhui University of Science and Technology, Huainan, Anhui 232001, China}
\author{W. Q. Li}
\affiliation{School of Electrical and Information Engineering, Anhui University of Science and Technology, Huainan, Anhui 232001, China}

\author{C. Y. Zheng}
\affiliation{Institute of Applied Physics and Computational Mathematics, Beijing, 100094, China}
\affiliation{\mbox{HEDPS, Center for Applied Physics and Technology, Peking University, Beijing 100871, China}}
\affiliation{\mbox{Collaborative Innovation Center of IFSA (CICIFSA), Shanghai Jiao Tong University, Shanghai 200240, China}}

\author{Z. J. Liu}
\affiliation{Institute of Applied Physics and Computational Mathematics, Beijing, 100094, China}
\affiliation{\mbox{HEDPS, Center for Applied Physics and Technology, Peking University, Beijing 100871, China}}

\author{L. H. Cao}
\affiliation{Institute of Applied Physics and Computational Mathematics, Beijing, 100094, China}
\affiliation{\mbox{HEDPS, Center for Applied Physics and Technology, Peking University, Beijing 100871, China}}
\affiliation{\mbox{Collaborative Innovation Center of IFSA (CICIFSA), Shanghai Jiao Tong University, Shanghai 200240, China}}

\author{C. Z. Xiao } %{$^{\dagger}$}
  \email{xiaocz@hnu.edu.cn}
  \affiliation{ School of Physics and Electronics, Hunan University, Changsha, 410082, China}
\affiliation{\mbox{Collaborative Innovation Center of IFSA (CICIFSA), Shanghai Jiao Tong University, Shanghai 200240, China}}
\date{\today}% It is always \today, today,
  \begin{abstract}

We present \scalebox{0.8}{\textsc{WEBS}} (WavE-Based Simulations), an efficient wave-based simulation model designed to investigate the dynamic interplay between cross-beam energy transfer (CBET) and stimulated Brillouin scattering (SBS) in laser-plasma systems. By employing a unified Schr$\ddot{o}$dinger-type envelope formulation for the laser and ion-acoustic waves, our model enables the use of a single, unconditionally stable Du Fort-Frankel numerical scheme, which maintains excellent long-term energy conservation even with coarse spatial grids. This approach not only achieves high computational efficiency validated against particle-in-cell simulations but also allows the selective activation or suppression of CBET and SBS processes, offering a clear diagnostic of their mutual coupling. Our simulations reveal that at high laser intensities, CBET and SBS reach a coupled steady state, leading to significant deviations from classical fluid theory predictions. Specifically, CBET gain is suppressed due to enhanced SBS reflectivity, while strong asymmetry in SBS reflectivity emerges between the interacting beams. These findings highlight regimes where the two instabilities strongly influence each other, providing critical insights for inertial confinement fusion research and offering a practical numerical tool for instability control and scenario design.
  \end{abstract}


  \maketitle

\section{Introduction}\label{introduction}

 In inertial confinement fusion (ICF), understanding and controlling laser-plasma instabilities remains a crucial challenge for achieving efficient energy coupling and symmetric implosion\cite{ICF1,ICF3,D_D,I_D,ICFn,Icfmre,Icfmre2,Icfmre3,Icfmre4}. Among these instabilities, cross-beam energy transfer (CBET)\cite{cbet5,cbet6,cbet7,cbet8,cbet9} and stimulated Brillouin scattering (SBS)\cite{sbs_m_l,sbs1,sbs2,sbs3,sbs4,sbs5,HaoL1}  represent fundamental processes that can significantly impact laser-plasma interactions in ICF scenarios\cite{press1,press2}. The crossed laser beams would induce the cross-beam energy transfer (CBET) when the phase match condition is satisfied. This instability is almost inevitable in ICF, however, people found that it is possible to tuning the implosion symmetry by CBET\cite{IV,cbet1,cbet2,cbet3,cbet10,cbet11,cbet12}.

Numerical simulation plays a crucial role in understanding laser plasma instabilities such as CBET and SBS, with various modeling approaches developed to balance physical accuracy and computational efficiency. Particle-in-cell (PIC) methods \cite{epoch,pic2,jiao}offer a kinetic description, capturing essential particle-scale effects like ion trapping and nonlinear Landau damping, but their high computational cost and intrinsic noise often limit the simulation of long-term, large-scale instability evolution. In contrast, fluid models provide a computationally efficient wave-wave interaction framework, enabling clear observation of linear growth from small perturbations and the investigation of instability saturation over extended spatial and temporal scales. High-fidelity models such as \scalebox{0.8}{\textsc{LPSE}}\cite{lpse,cbet13} solve the full-wave electromagnetic equations and serve as a benchmark for accuracy, especially in regions where ray-based\cite{ray1,ray2} approximations fail. To enhance computational efficiency while retaining wave resolved physics, contemporary approaches such as the Hamiltonian based \scalebox{0.8}{\textsc{BEAM}} \cite{jiaxb} adopt symplectic time integration techniques, thereby ensuring numerical stability and strict energy conservation over long simulation intervals. Meanwhile, spectral-decomposition methods exemplified by \scalebox{0.8}{\textsc{HARMONY}} \cite{har,raj} efficiently capture the saturation of SBS through ion acoustic harmonic generation, enabling large scale simulations at manageable computational cost.

Notwithstanding these advancements, a gap remains for a computationally efficient, flexible, and numerically robust wave-based framework that can simulate the coupled and time-resolved evolution of CBET and SBS in multidimensional geometries over experimentally relevant durations. In this paper,  we introduce \scalebox{0.8}{\textsc{WEBS}} (WavE-Based Simulations), a parallel simulation code designed specifically for studying the interplay between these instabilities. Departing from existing methodologies, \scalebox{0.8}{\textsc{WEBS}} employs a unified Schr$\ddot{o}$dinger-type envelope representation for both electromagnetic and ion acoustic waves, which permits all coupled equations to be integrated using a single, unconditionally stable Du Fort-Frankel scheme. This consolidated numerical strategy confers several notable benefits: (1) it sustains excellent long-term energy conservation even with relatively coarse spatial discretizations, substantially lowering computational overhead compared with conventional explicit solvers; (2) it affords straightforward selective activation or suppression of CBET and SBS processes, thereby facilitating the unambiguous diagnosis of their mutual coupling; and (3) it delivers results that align closely with both PIC benchmarks and well established fluid theory predictions in the appropriate limits. Leveraging \scalebox{0.8}{\textsc{WEBS}}, we systematically explore the coupled steady state attained by CBET and SBS, delineate parameter ranges in which their interaction appreciably alters energy transfer dynamics, and quantify departures from classical gain estimates under high intensity conditions. It is found that after the hybrid system reaches a steady state, the gain of CBET may deviate from the predictions of fluid-based CBET theory, which is attributed to the enhanced SBS reflectivity following laser energy transfer. Furthermore, we demonstrate that under conditions of high laser intensity and low Landau damping, the interplay between CBET and SBS cannot be neglected.

  This paper is structured in the following ways. Firstly, in Sec.~\ref{algorithm}, we describe the fluid model of CBET and SBS with coupling equations in Schr$\ddot{\rm o}$dinger equation's form.  We simulate CBET by \scalebox{0.8}{\textsc{WEBS}} in homogeneous plasma in Sec.~\ref{cbetwebs}. We also simulation SBS processes by \scalebox{0.8}{\textsc{WEBS}} and comparing with theory in   Sec.~\ref{SBS model}. Then, the interaction between CBET and SBS shown in Sec.~\ref{SBS cbet}. At last, the conclusion  and discussion are shown in  Sec.~\ref{conclusion}.

\section{The fundamental equations and algorithms for WEBS code}\label{algorithm}

Cross-beam energy transfer is a prevalent instability in inertial confinement fusion (ICF), describing the process where intersecting laser beams exchange energy through ion acoustic waves (IAW) in plasmas. In the fundamental CBET model, two laser beams intersect at a specific angle, with one beam acting as the pump laser that loses energy, while the other serves as the seed laser that gains energy. This energy transfer is facilitated by ion acoustic waves, and the interaction among these three waves can be mathematically described by a system of coupled equations\cite{wangqing2,wangqing1,rosenbluth1,rosenbluth2},

\begin{equation} \label{cbet1}
 \begin{split}
    &(\partial^{2}_{t}-c^{2}\nabla^{2}+\omega_{pe}^{2}  ) A_{0} = -\frac{4\pi e^{2}}{m_{e}} \tilde{n}_{e} A_{1},\\
    &(\partial^{2}_{t}-c^{2}\nabla^{2}+\omega_{pe}^{2}  )A_{1} = -\frac{4\pi e^{2}}{m_{e}} \tilde{n}_{e} A_{0},\\
    &\big[(\partial_{t}+V\cdot\nabla)^{2}+2\nu(\partial_{t}+V\cdot\nabla)-C_{s}^{2}\nabla^{2}\big]\tilde{n}_{e}=\\
    &\frac{Zn_{0}e^{2}}{m_{e}m_{i}c^{2}}\nabla^{2}(A_{0}\cdot A_{1}),
  \end{split}
\end{equation} where $A_{0}$ $A_{1}$ are the vector potential of pump laser and seed laser, respectively. $\tilde{n}_{e}$ is the plasma density perturbation of ion acoustic wave. $c$ is the velocity of light in vacuum,  $n_{0}$ and $\omega_{pe}$ are the  electron density and plasma frequency, $m_{e}$ and $m_{i}$ are the mass of electron and ion, respectively, $Z$ is the charge state of ion, and $C_{s}$ is the ion-sound velocity, $\nu$ is the landau damping of IAW.

  The intensities of laser beams in ICF are in the range of $10^{14}\sim10^{15}$ $\rm{W/cm^{2}}$, so, the characteristic time of CBET can be obtained by $\tau = 1/\gamma_{0}$, which is around thousands of times of pump laser's period, where $\gamma_{0}$ is the growth rate of CBET in homogenous plasma. Thus, the three waves of SBS can be expressed as the product of a slowly-changing envelope in time and a rapidly-oscillating phase,
\begin{equation} \label{cbet_4}
\begin{split}
&A_{0}(\vec{r},t) = \frac{1}{2}\tilde{A}_{0}(\vec{r},t)e^{-i\omega_{0}t+i\vec{k}_{0}\cdot \vec{r}} + c.c.\\
&A_{1}(\vec{r},t) = \frac{1}{2}\tilde{A}_{1}(\vec{r},t)e^{-i\omega_{1}t+i\vec{k}_{1}\cdot \vec{r}} + c.c.\\
&\tilde{ n}_{e}(\vec{r},t) = \frac{1}{2}\delta \tilde{n}_{e}(\vec{r},t)e^{-i\omega_{2}t+i\vec{k}_{2}\cdot \vec{r}+i\int_{0}^{t}\delta\omega dt} + c.c.
\end{split}
\end{equation} where $\omega_{0}$ is the frequency of pump laser, $\omega_{1}$ is the frequency of seed laser and $\omega_{2}$ is the frequency of IAW.  $e^{i\int_{0}^{t}\delta\omega dt}$ is the phase-mismatch induced by the nonlinear frequency shift of IAW.

Substituting Eqs.~(\ref{cbet_4}) to Eqs.~(\ref{cbet1}), Then,we obtain the coupling equations with Schr$\ddot{\rm o}$dinger equation's from by neglecting the second derivative of time,
\begin{equation} \label{cbet5}
    \partial_{t}a_{0} = \frac{i}{2}\big[\nabla^{2}+(1-n_{0})\big]a_{0}-\frac{i}{4}\tilde{n}_{e}a_{1},
\end{equation}
\begin{equation} \label{cbet6}
    \partial_{t}a_{1} =\frac{i}{2}\frac{\omega_{0}}{\omega_{1}}\bigg[\nabla^{2}+(\frac{\omega_{1}^{2}}{\omega_{0}^{2}}-n_{0})\bigg]a_{1}-\frac{i}{4}\frac{\omega_{0}}{\omega_{1}}\tilde{n}_{e}^{\ast}a_{0},
\end{equation}

\begin{equation} \label{cbet7}
\begin{split}
 \partial_{t}\tilde{n}_{e}&=\frac{i}{2} \frac{\omega_{0}}{\omega_{20}}\bigg[\frac{C_{s}^{2}}{c^{2}}\nabla^{2}+\frac{\omega_{2}^{2}}{\omega_{0}^{2}}-\frac{V_{x}^{2}}{c^{2}}\partial_{x}^{2}-\frac{V_{y}^{2}}{c^{2}}\partial_{y}^{2}  \bigg]\tilde{n}_{e}\\
 &-\frac{\omega_{2}}{\omega_{20}}\bigg(\frac{V_{x}}{c}\partial_{x}+\frac{V_{y}}{c}\partial_{y}\bigg)\tilde{n}_{e} - (\nu +i\delta\omega)\tilde{n}_{e}\\
 & +\frac{i}{4}\frac{Zn_{0}m_{e}\omega_{0}}{m_{i}\omega_{20}}\nabla^{2}(a_{1}^{\ast}\cdot a_{0}) ,
 \end{split}
\end{equation}where $a_{0} = \frac{eA_{0}}{m_{e}c^{2}}$, $a_{1} = \frac{eA_{1}}{m_{e}c^{2}}$ are normalized  amplitude of pump and seed, respectively,  $\tilde{n}_{e}$ and $n_{0}$ have been normalized to the critical density of pump laser $n_{c}$, $V_{x}$ and $V_{y}$ represent the plasma flow velocity components along each axis, where their positive directions correspond to the projections of $C_{s}$ onto the $X$- and $Y$-directions, respectively. The spatial and temporal coordinates are normalized to $c/\omega_{0}$ and $\omega_{0}^{-1}$, respectively.

 It should be noted that $\omega_{20}$ represents the IAW frequency when the plasma flow velocity vanishes. The Landau damping and nonlinear frequency shift of IAWs can be expressed as:
\begin{equation} \label{iaw_ld}
 \begin{split}
 &\nu = \frac{\omega_{20}}{\omega_{0}}\bigg[ \sqrt{\frac{Zm_{e}}{m_{i}}}+\bigg(\big(\frac{ZT_{e}}{T_{i}}\big)^{\frac{3}{2}} e^{-\frac{(\vec{C_{s}}-\vec{V})^{2}}{2v_{i}^{2}}}\bigg)\bigg]\bigg/(1+k^{2}\lambda_{De}^{2})^{\frac{3}{2}},\\
& \delta\omega = - \frac{1}{\sqrt{2\pi}}\bigg[ \alpha_{i}\sqrt{\frac{ZT_{e}}{T_{i}   }} ( \varsigma^4 - \varsigma^2)e^{-\frac{\varsigma^2}{2}} - \alpha_{e}\bigg]|\tilde{n}_{e}/n_{0}|^{\frac{1}{2}},
 \end{split}
\end{equation}where $\lambda_{De}$ is the Debye length of plasma. We note that the \scalebox{0.8}{\textsc{WEBS}} code currently accounts for kinetic nonlinear frequency shifts, but does not incorporate fluid nonlinear frequency shifts. The parameters $\alpha_{e} = 0.544$, which stands for the contributions to  the frequency shift from electrons and ions\cite{shift},where $\varsigma = C_{s}/v_{thi}$, $v_{thi} = \sqrt{T_{i}/m_{i}}$.

 %When the plasma flow velocity balances the ion sound speed, i.e., when $C_{s} = \sqrt{V_{x}^{2} + V_{y}^{2}}$, Eq.~(\ref{cbet7}) reduces to
%\begin{equation} \label{cbet7_1}
%  \partial_{t}\tilde{n}_{e}=-(\nu +i\delta\omega)\tilde{n}_{e}-\frac{\omega_{2}}{\omega_{20}}\bigg(\frac{V_{x}}{c}\partial_{x}+\frac{V_{y}}{c}\partial_{y}\bigg)\tilde{n}_{e}+\frac{i}{4}\frac{Zn_{0}m_{e}\omega_{0}}{m_{i}\omega_{20}}\nabla^{2}(a_{1}^{\ast}\cdot a_{0}),
%\end{equation}indicating that the ion-acoustic wave (IAW) transitions to a standing wave.

In the absence of plasma flow ($V = 0$), Eq.~(\ref{cbet7}) simplifies to
\begin{equation} \label{cbet7_2}
 \begin{split}
  \partial_{t}\tilde{n}_{e}&=\frac{i}{2} \frac{\omega_{0}}{\omega_{20}}\bigg[\frac{C_{s}^{2}}{c^{2}}\nabla^{2}+\frac{\omega_{20}^{2}}{\omega_{0}^{2}}\bigg]\tilde{n}_{e}-(\nu +i\delta\omega)\tilde{n}_{e}\\
  &+\frac{i}{4}\frac{Zn_{0}m_{e}\omega_{0}}{m_{i}\omega_{20}}\nabla^{2}(a_{1}^{\ast}\cdot a_{0}),
\end{split}
\end{equation}

We present \scalebox{0.8}{\textsc{WEBS}} (WavE-Based Simulation), a parallel computational framework designed to solve coupled laser-plasma instability equations. The code implements the Du Fort-Frankel scheme to numerically integrate Eqs.~(\ref{cbet5}-\ref{cbet7}) and currently supports the simulation of instabilities such as CBET and SBS. Future releases are planned to extend this capability to include Stimulated Raman Scattering (SRS) and Two-Plasmon Decay (TPD)\cite{TPD}.

The Du Fort-Frankel scheme, originally proposed in 1953~\cite{DFF} for second-order parabolic differential equations, has been successfully applied to nonlinear problems~\cite{DFF2,DFF3}. This scheme offers two key advantages: unconditional stability~\cite{DFF5} and second-order accuracy with truncation error $O(\Delta t^2/h^2)$\cite{DFF4}, where $\Delta t$ and $h$ represent temporal and spatial grid sizes respectively. Similar to leapfrog schemes and symplectic algorithms~\cite{jiaxb,xin1,xin2}, our \scalebox{0.8}{\textsc{WEBS}} code maintains excellent long-term energy conservation properties, making it particularly suitable for CBET and SBS simulations where energy balance is crucial.

  In \scalebox{0.8}{\textsc{WEBS}} , the full time steps of waves are denoted by $n-1, n, n+1$, and the half time steps of waves are denoted by $n-1/2,n+1/2,n+3/2$. space is discretized by $\Delta x$ and $\Delta y$ at two dimensions, for simplicity, we use $\Delta x = \Delta y = h$, space points are denoted by $(i,j)$. The Du Fort-Frankel scheme has three folds in time, which mean that the value of $u^{n+1}$ requires $u^{n}$ and $u^{n+\frac{1}{2}}$.  For example, The Du Fort-Frankel scheme for pump laser is
  \begin{equation} \label{DF1}
   \begin{split}
    a&^{n+1}_{0,(i,j)} = \frac{1-2\xi_{0}}{1+2\xi_{0}}a^{n}_{0,(i,j)}+\frac{\xi_{0}}{1+2\xi_{0}}\bigg[a^{n+\frac{1}{2}}_{0,(i-1,j)} \\
                       &+a^{n+\frac{1}{2}}_{0,(i+1,j)}+a^{n+\frac{1}{2}}_{0,(i,j-1)}+a^{n+\frac{1}{2}}_{0,(i,j+1)}\bigg]\\
                       &+\frac{\Delta t}{1+2\xi_{0}}\bigg[\frac{i}{2}(1-n_{0})a^{n+\frac{1}{2}}_{0,(i,j)}-\frac{i}{4}\tilde{n}_{e,(i,j)}^{n+\frac{1}{2}}a_{1,(i,j)}^{n+\frac{1}{2}}\bigg],
   \end{split}
  \end{equation} where $\xi_{0} = \frac{i\Delta t}{2h^{2}}$. Similarly, the  Du Fort-Frankel scheme for seed laser is
  \begin{equation} \label{DF2}
   \begin{split}
    a&^{n+1}_{1,(i,j)} = \frac{1-2\xi_{1}}{1+2\xi_{1}}a^{n}_{1,(i,j)}+\frac{\xi_{1}}{1+2\xi_{1}}\bigg[a^{n+\frac{1}{2}}_{1,(i-1,j)} \\
                       &+a^{n+\frac{1}{2}}_{1,(i+1,j)}+a^{n+\frac{1}{2}}_{1,(i,j-1)}+a^{n+\frac{1}{2}}_{1,(i,j+1)}\bigg] +\\
                       &\frac{\Delta t}{1+2\xi_{1}}\bigg[\frac{i}{2}\bigg(\frac{\omega_{1}^{2}}{\omega_{0}^{2}}-n_{0}\bigg)a^{n+\frac{1}{2}}_{1,(i,j)}- \frac{i}{4}\frac{\omega_{0}}{\omega_{1}}\tilde{n}_{e,(i,j)}^{{\ast}n+\frac{1}{2}}a_{0,(i,j)}^{n+\frac{1}{2}}\bigg],
   \end{split}
  \end{equation}where $\xi_{1} = \frac{i\omega_{0}\Delta t}{2\omega_{1}h^{2}}$. The Du Fort-Frankel scheme for IAW is a little more complicated, which is
  \begin{equation} \label{DF3}
   \begin{split}
   &\tilde{n}_{e,(i,j)}^{n+1}=\frac{1-C_{1}-C_{2}}{1+C_{1}+C_{2}}\tilde{n}_{e,(i,j)}^{n}+\frac{C_{1}+C_{4}}{1+C_{1}+C_{2}}\tilde{n}_{e,(i+1,j)}^{n+\frac{1}{2}}\\
                           &+\frac{C_{1}-C_{4}}{1+C_{1}+C_{2}}\tilde{n}_{e,(i-1,j)}^{n+\frac{1}{2}}+\frac{C_{2}+C_{5}}{1+C_{1}+C_{2}}\tilde{n}_{e,(i,j+1)}^{n+\frac{1}{2}}\\
                           &+\frac{C_{2}-C_{5}}{1+C_{1}+C_{2}}\tilde{n}_{e,(i,j-1)}^{n+\frac{1}{2}}+\frac{C_{3}+C_{7}+C_{8}}{1+C_{1}+C_{2}}\tilde{n}_{e,(i,j)}^{n+\frac{1}{2}}\\
                           &+\frac{C_{6}}{1+C_{1}+C_{2}} \bigg[\frac{4\big(T_{(i-1,j)}+T_{(i+1,j)}+T_{(i,j-1)}+T_{(i,j+1)}\big)}{h^{2}}\\
                           &+\frac{T_{(i-1,j-1)}+T_{(i+1,j+1)}+T_{(i-1,j-1)}+T_{(i+1,j+1)}}{h^{2}}-\frac{20T_{(i,j)}}{h^{2}}\bigg],
  \end{split}
  \end{equation} where $T_{(i,j)} = a_{1,(i,j)}^{\ast n+\frac{1}{2}}\cdot a_{0,(i,j)}^{n+\frac{1}{2}}$, the coefficients $C_{1}$$\thicksim$$C_{8}$ are
  \begin{equation} \label{DF4}
   \begin{split}
    &C_{1} = \frac{i}{2}\frac{\omega_{0}\Delta t}{\omega_{20}h^{2}}\frac{C_{s}^{2}-V_{x}^{2}}{c^{2}}, C_{2} = \frac{i}{2}\frac{\omega_{0}\Delta t}{\omega_{20}h^{2}}\frac{C_{s}^{2}-V_{y}^{2}}{c^{2}}\\
    &C_{3} = \frac{i}{2}\frac{\omega_{2}^{2}\Delta t}{\omega_{0}\omega_{20}},    C_{4} = -\frac{\omega_{2}\Delta t}{2\omega_{20}h}\frac{V_{x}}{c}\\
    &C_{5} = -\frac{\omega_{2}\Delta t}{2\omega_{20}h}\frac{V_{y}}{c},     C_{6} = \frac{i}{4}\frac{Zn_{0}m_{e}\omega_{0}\Delta t}{m_{i}\omega_{20}}\\
    &C_{7} = -\nu \Delta t, C_{8} = -i \Delta t \delta\omega.
   \end{split}
  \end{equation}

 It can be observed that the wave field update at position $(i,j)$ only requires data from its immediate neighborhood, demonstrating the algorithm's natural parallelism.

\section{ CBET simulations by WEBS in homogeneous plasmas  }\label{cbetwebs}

 \begin{figure}[htbp]
    \begin{center}
      \includegraphics[width=0.52\textwidth,clip,angle=0]{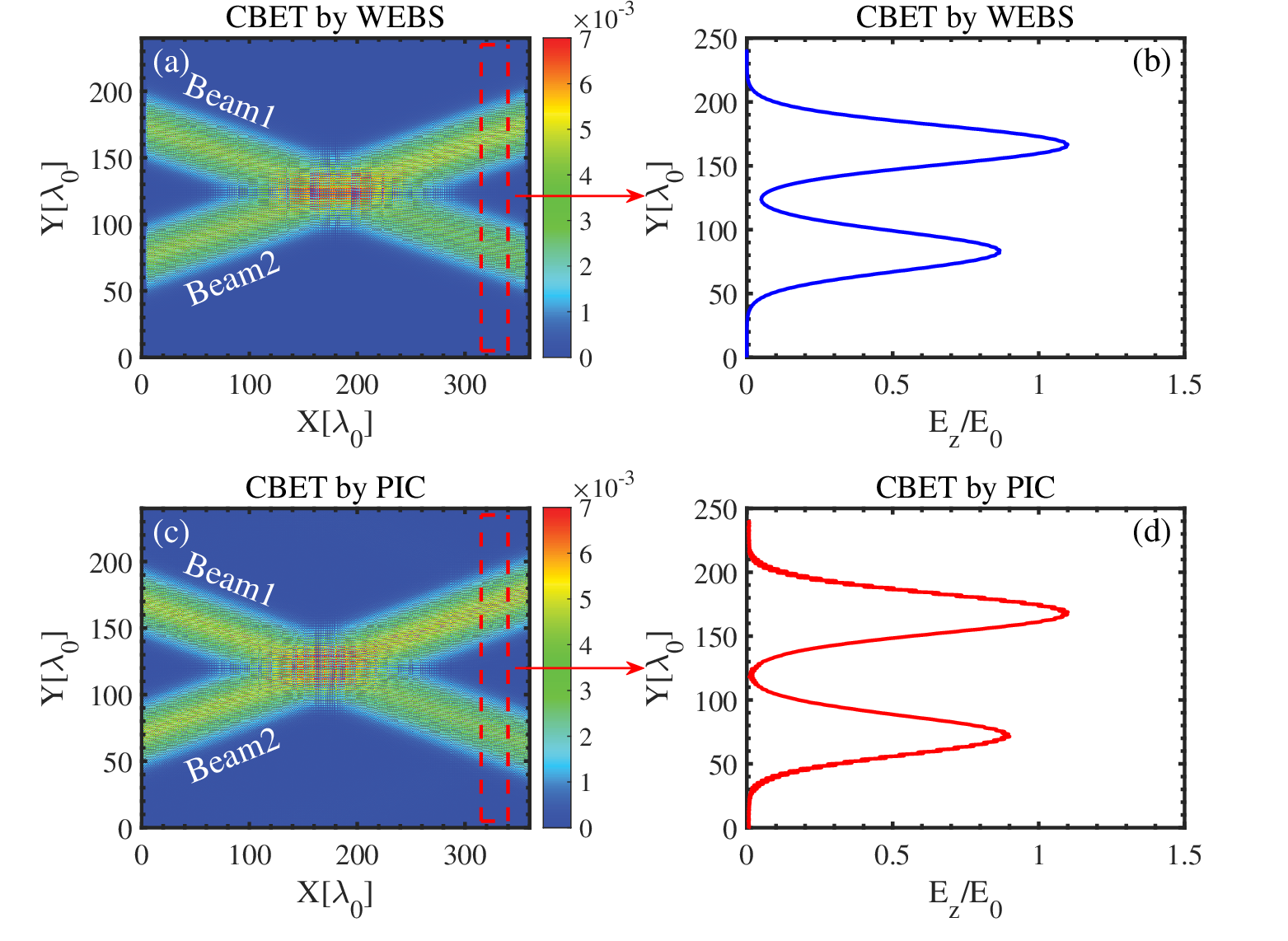}\vspace{-10pt}
      % Here is how to import EPS art [width=20mm,height=10mm][width=9cm,trim=0.5cm 21cm 0 2cm]
      \caption{\label{tw_pic} The comparison between WEBS simulation results and PIC simulation results.(a) The amplitude of laser at $t = 6.6 \rm{ps}$ by WEBS simulation. (b) the normalized amplitude of two lasers after CBET, the results are obtained by the integral of $x$
      in the red rectangle of (a). (c) The amplitude of laser at $t = 6.6 \rm{ps}$ by PIC simulation. (d) the normalized amplitude of two lasers after CBET, the results are obtained by the integral of $x$
      in the red rectangle of (c). }
    \end{center}
  \end{figure}

\begin{figure}[htbp]
    \begin{center}
      \includegraphics[width=0.5\textwidth,clip,angle=0]{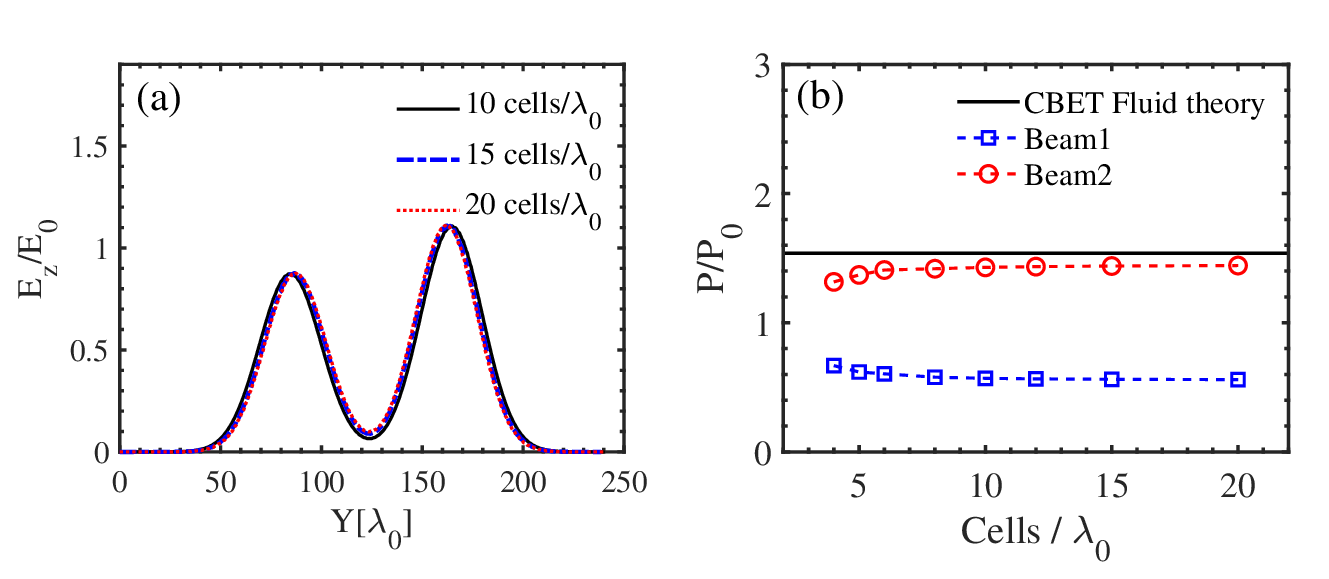}\vspace{-10pt}
      % Here is how to import EPS art [width=20mm,height=10mm][width=9cm,trim=0.5cm 21cm 0 2cm]
      \caption{\label{mesh} The results of WEBS for different spatial mesh. The parameters of lasers and plasma are same to that in  Fig~\ref{tw_pic}. The black line is the amplitude of two laser at $t = 6.6\rm{ps}$ by spatial meshes $10~\mathrm{cells}/\lambda_0$. The blue dashed line is the amplitude of two laser at $t = 6.6\rm{ps}$ by spatial meshes $15~\mathrm{cells}/\lambda_0$. The red dot line is the amplitude of two laser at $t = 6.6\rm{ps}$ by spatial meshes $20~\mathrm{cells}/\lambda_0$. }
    \end{center}
  \end{figure}

 A comparison is presented between the CBET results predicted by the \scalebox{0.8}{\textsc{WEBS}} model and those from PIC simulations using the \textsc{epoch} code~\cite{epoch}. We model a two-dimensional (2D) CBET scenario in which two Gaussian laser beams intersect at a $31^\circ$ angle after entering from the left boundary of the simulation domain. Both beams share a peak intensity of $3\times10^{14}~\mathrm{W/cm^{2}}$ and a lateral full-width-at-half-maximum (FWHM) of $33.3\lambda_0$, where $\lambda_0 = 0.351~\mathrm{\mu m}$ is the pump laser wavelength. The simulation box measures $L_x = 360\lambda_0$ by $L_y = 240\lambda_0$. It contains a uniform helium plasma ($Z=2$, $m_i=7344m_e$) at a density of $0.05n_c$, with $n_c$ denoting the critical density. The electron and ion temperatures are fixed at $T_e=1~\mathrm{keV}$ and $T_i=0.2~\mathrm{keV}$, respectively. To ensure resonant CBET excitation, the seed laser wavelength is set to $\lambda_1=0.3512~\mathrm{\mu m}$

The PIC simulations utilize a spatial resolution of 10 cells per $\lambda_0$ and 100 particles per cell (PPC), with all laser and plasma parameters aligned with the \scalebox{0.8}{\textsc{WEBS}} configuration; both models assume zero initial plasma flow velocity.

Figures~\ref{tw_pic}(a) and (c) display the normalized laser amplitudes from the \scalebox{0.8}{\textsc{WEBS}} and PIC simulations, revealing consistent features between the two methods. The attenuation of the pump laser within the overlap region clearly signifies energy transfer via CBET. To quantify this transfer, we analyze the laser amplitude within the red-dashed areas in Figs.~\ref{tw_pic}(a) and (c), with the results plotted in panels (b) and (d). The \scalebox{0.8}{\textsc{WEBS}} simulations (at $10~\mathrm{cells}/\lambda_0$) reproduce the PIC results with excellent agreement, particularly in capturing the maximum amplitudes of both the pump and seed lasers.

 \begin{figure}[htbp]
    \begin{center}
      \includegraphics[width=0.5\textwidth,clip,angle=0]{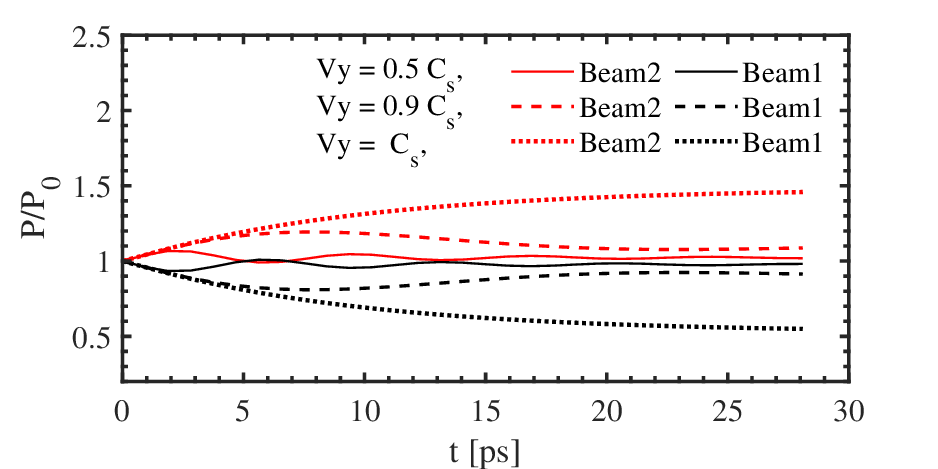}\vspace{-10pt}
      % Here is how to import EPS art [width=20mm,height=10mm][width=9cm,trim=0.5cm 21cm 0 2cm]
      \caption{\label{meshvy} The results of WEBS for different flowing velocities of plasmas. The parameters of lasers are same to that in  Fig~\ref{tw_pic}. The black curves represents the temporal evolution of the pump laser power, while the red curves corresponds to that of the seed laser power. Three different flow velocity cases are considered, CBET operates in the resonant regime when $V_{y} = C_{s}$.  }
    \end{center}
  \end{figure}

A further investigation into the sensitivity of WEBS results to spatial resolution is presented in Fig.~\ref{mesh}. In panel (a), the results for resolutions of 10, 15, and 20 cells/$\lambda_0$ (shown as black solid, blue dashed, and red dotted curves, respectively) are in close agreement. This indicates that the physical results are robust and that computational costs can be reduced by employing coarser spatial meshes. Panel (b) displays the convergence of CBET simulation results with increasing grid density. It can be seen that the solution converges for resolutions beyond 10 cells/$\lambda_0$. The marginal accuracy gain from higher grid densities (15-20 $\mathrm{cells}/\lambda_0$) is outweighed by their significantly greater computational requirements. In all cases, the CFL condition is strictly enforced by setting $\Delta t = 0.2\Delta h$.

Then, we compare our simulation results with  CBET fluid theory, The fluid theory of CBET is frequently used to study the CBET processes in ICF\cite{IV}, so, we reuse the fluid theory of CBET to obtain the gain of CBET. In our model, the intensity growth of Beam2 along its path $\ell$ is,
\begin{equation}\label{gain7}
 I_{2}^{out} = I_{2}^{in} \rm{exp}\bigg[\int L^{-1}_{CBET} d\ell \bigg],
\end{equation}   where $I_{2}^{in}$ is the original intensity of Beam2, $I_{2}^{out}$ is the intensity of Beam2 after the CBET. $L_{\rm{CBET}}$ is the scale-length of energy gain by CBET, $L^{-1}_{\rm{CBET}}$ can be obtained by
\begin{equation}\label{gain8}
 L^{-1}_{\rm{CBET}} = \frac{k_{0}}{4}\frac{n_{0}}{n_{c}-n_{0}}\bigg( \frac{\bar{v}_{osc}}{v_{e}}\bigg)^{2}\bigg[ (1+3T_{i}/ZT_{e})\frac{\nu_{2}}{\omega_{20}}\bigg]P(\eta),
\end{equation} where $\nu_{2}$ is the Landau damping of IAW, $\bar{v}_{osc}$ is the average normalized amplitude over the  width of Gaussian lasers, $d = 90 \lambda_{0}$, and $P(\eta)$ is defined by
\begin{equation}\label{gain9}
 P(\eta) = \frac{(\nu_{2}/\omega_{20})^{2}\eta}{(\eta^{2}-1)^{2}+(\nu_{2}/\omega_{20})^{2}\eta^{2}},
\end{equation} the $\eta$ is the modifying coefficient of IAW frequency because of the flowing velocity of plasma, $\eta = \vec{k_{2}}\cdot \vec{V}/\omega_{20}+\omega_{2}/\omega_{20}$.
From Eq.~(\ref{gain7}), the energy gain of CBET without considering pump depletion can be expressed as
\begin{equation}\label{gain10}
 G_{0} = \int L^{-1}_{\rm{CBET}} d\ell ,
\end{equation} which is the spatial gain of Beam2 along the path $\ell$. In the homogenous plasma, $L^{-1}_{CBET}$ is a constant, then the Eq.~(\ref{gain10}) can be reduced to
 \begin{equation}\label{gain10_1}
 G_{0} = L^{-1}_{\rm{CBET}} \cdot L_{c} ,
\end{equation} where $L_{c}=d/sin\theta$ is the reduction of interaction length of CBET.

 However, this equation dose not consider the influence of pump depletion, we should use Tang formula\cite{tang} to obtain the final spatial gain of CBET with considering pump depletion\cite{cbet9},
\begin{equation}\label{gain11}
 G_{\rm{CBET}} = \rm{ln} \big(I_{2}^{out}/I_{2}^{in}\big) = \rm{ln}\bigg[\frac{(1+\beta)e^{G_{0}(1+\beta)}}{1+\beta e^{G_{0}(1+\beta)}}\bigg],
\end{equation} where $\beta = I_{2}^{in}/I_{0}$, is the seed level for CBET, When the $\beta$ is small enough, the $G_{\rm{CBET}}$ will be equal to $G_{0}$.   In our model, we assume that two crossed laser beams have the same intensity, so, $\beta = 1$. The black solid line in Fig.~\ref{mesh} represents the power of Beam 2 relative to its initial value. This result is derived from the fluid theory of CBET, specifically based on Eq.~(\ref{gain11}), We observe that the simulation results converge toward the theoretical values as the number of cells increases.

  The capability of the \scalebox{0.8}{\textsc{WEBS}} code to model the effect of plasma flow velocity on CBET is investigated. A scenario with two intersecting Gaussian laser beams of identical frequency is considered, propagating in a uniform plasma with a flow velocity $V_y$ along the $Y$-direction. It is demonstrated that when $V_y = C_s$, the flow effectively cancels the ion acoustic wave, fulfilling the three-wave resonance condition and resulting in maximum energy transfer, as indicated by the black and red dotted curves in Fig.~\ref{meshvy}. When the flow velocity is reduced, the system moves away from resonance, leading to diminished CBET. This is confirmed by the results for $V_y = 0.9C_s$ and $V_y = 0.5C_s$ in Fig.~\ref{meshvy}.

\section{SBS simulations by WEBS in homogeneous and inhomogeneous plasmas}\label{SBS model}

In addition to simulating the CBET process in inertial confinement fusion, the \scalebox{0.8}{\textsc{WEBS}} code is also capable of modeling the SBS process. In this section, we will employ the WEBS code to simulate SBS in both uniform plasma and non-uniform plasma with flow velocity.

We begin by simulating SBS in a uniform plasma using the WEBS code. The simulation domain has a longitudinal length of $180\lambda_0$ and a transverse length of $120\lambda_0$. The plasma parameters are identical to those used in the previous CBET simulation, with no flow velocity. A Gaussian pump laser with a peak intensity of $10^{15}~\mathrm{W/cm^2}$ and a full width at half maximum (FWHM) of $20\lambda_0$ is incident from the left boundary. Simultaneously, a seed laser, with an intensity of $10^{-8}$ times that of the pump laser, is injected from the right boundary.

\begin{figure}[htbp]
    \begin{center}
      \includegraphics[width=0.5\textwidth,clip,angle=0]{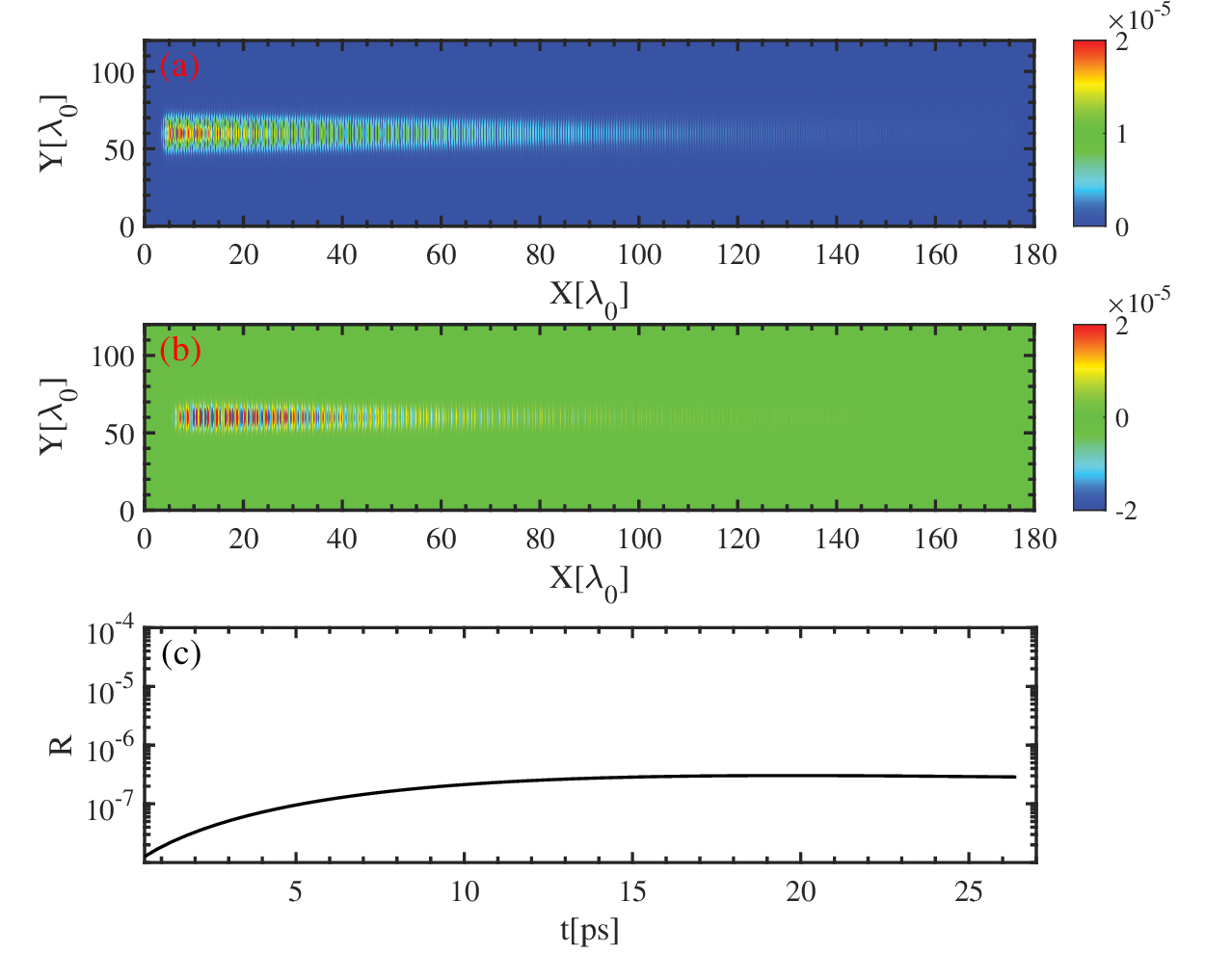}\vspace{-10pt}
      % Here is how to import EPS art [width=20mm,height=10mm][width=9cm,trim=0.5cm 21cm 0 2cm]
      \caption{\label{hom_fig} The WEBS simulation results of SBS in homogeneous plasma. (a) Spatial distribution of the seed laser at $t=26.4$ ps.(b) Corresponding spatial distribution of the ion-acoustic waves. (c) Temporal evolution of the SBS reflectivity.}
    \end{center}
  \end{figure}
  
Due to the presence of ion-acoustic wave Landau damping, the SBS in this simulation exhibits convective instability. Both the seed laser light and the ion-acoustic waves grow over time before saturating. Fig.~\ref{hom_fig}(a) and 4(b) show the saturated seed light and ion-acoustic waves, respectively, in the steady state. It can be observed that the intensities of the reflected light and the ion-acoustic waves are highest in the region where the pump laser is incident. This is because the phase-matching condition for SBS is satisfied throughout the entire interaction region. Fig.~\ref{hom_fig}(c) illustrates the temporal evolution of SBS reflectivity. As shown, the reflectivity initially increases with time and subsequently saturates at approximately $2\times10^{-7}$, a clear signature of convective instability.

Convective SBS in an inhomogeneous flowing plasmas are simulated using the \scalebox{0.8}{\textsc{WEBS}} code. The simulation domain spans $360\lambda_0$ in the longitudinal direction and $120\lambda_0$ in the transverse direction. A linearly varying flow velocity profile is imposed: $V(x) = C_s \bigl( -1.45 + 0.9\frac{x}{x_{\mathrm{end}}} \bigr)$, where $x_{\mathrm{end}}$ is the total plasma length. At the longitudinal center ($x = x_{\mathrm{end}}/2$), the flow velocity and electron density are set to $-C_s$ and $0.05n_c$, respectively. To satisfy mass conservation, the corresponding density profile is derived as $n_e(x) = -0.05C_s / V(x)$. From this profile, the characteristic scale lengths for density and velocity are both found to be $L_n = L_V = 400\lambda_0$. The plasma temperature remains the same as in the homogeneous plasma simulations described earlier, while the intensity of the seed laser is set to $10^{-6}$ $I_{\text{pump}}$. The pump and seed lasers share the same frequency, so the SBS resonance occurs at the center of the simulation box, where the plasma flow velocity exactly compensates the ion acoustic speed.

\begin{figure}[htbp]
    \begin{center}
      \includegraphics[width=0.5\textwidth,clip,angle=0]{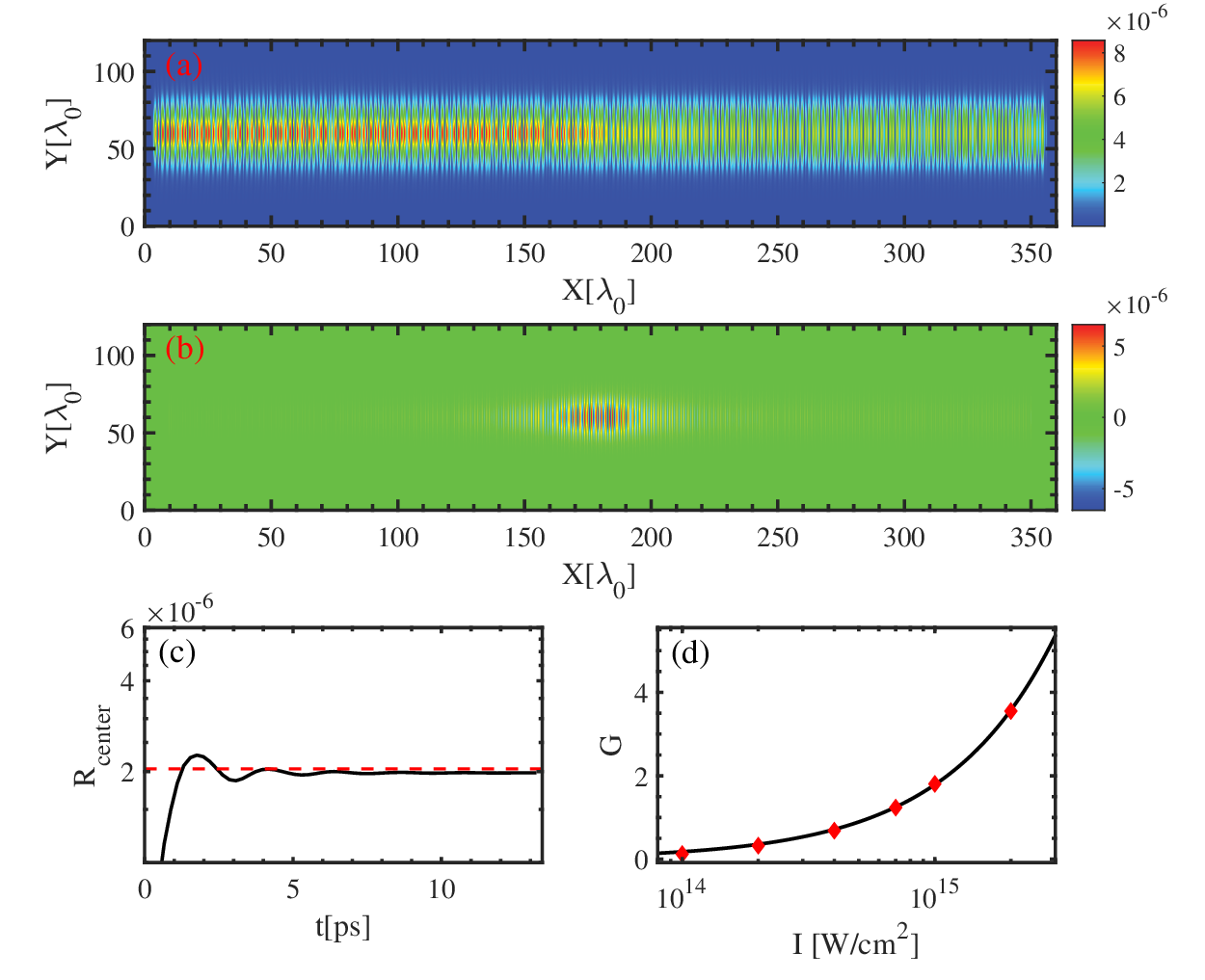}\vspace{-10pt}
      % Here is how to import EPS art [width=20mm,height=10mm][width=9cm,trim=0.5cm 21cm 0 2cm]
      \caption{\label{inhom_fig} The WEBS simulation results in inhomogeneous plasmas. (a) and (b) are the Spatial distribution of seed laser and IAWs at $t = 13.2$ $ps$, respectively when the peak intensity of pump is $4\times 10^{14}$ $\rm{W/cm^2}$. (c) black line is the reflectivity of SBS at the center of pump laser, red dashed line is the reflectivity by Rosenbluth theory. (d) The gains of SBS for different pump intensities, black line is obtained by Eq.~(\ref{sbs_ros}), red diamonds are the results from WEBS simulations.    }
    \end{center}
  \end{figure}

As shown in Fig. \ref{inhom_fig} (a) and (b), the spatial distributions of the seed laser and the ion acoustic waves are presented for a laser intensity of $4\times10^{14}~\mathrm{W/cm^2}$. In Fig. \ref{inhom_fig} (a), the seed laser is amplified only within the resonant region (i.e., the center of the simulation domain). Correspondingly, Fig. \ref{inhom_fig} (b) shows that the ion acoustic waves exhibit the highest intensity in the same resonant area. To compare with the convective gain of SBS proposed by Rosenbluth, the reflectivity at the center of the Gaussian beam at $Y$ direction, denoted as $R_{center}$, is collected, which corresponds to the one-dimensional SBS reflectivity under simplified conditions. In Fig.~\ref{inhom_fig} (c), the black curve displays the reflectivity at the center of the Gaussian laser beam, saturating at approximately $6 \times 10^{-6}$. The red dashed line represents the Rosenbluth gain for SBS backscattering\cite{sbs5,rosenbluth1,rosenbluth2}, which is obtained by
\begin{equation}\label{sbs_ros}
 R = 10^{-6}e^{G_{ros}}, G_{ros} = \frac{2\pi k_{iaw} a_0^{2} \omega_{pi}^{2} L_{V}}{16 k_{iawx} V_0 k_{sr} C_{s}cos\theta }
\end{equation}, where $\omega_{pi}$ is the ion frequency of plasma,  $k_{\text{iaw}}$ is the wavenumber of IAWs, $k_{\text{iawx}}$ is the longitudinal component of $k_{\text{iaw}}$, $a_0$ is the normalized intensity of the pump laser, $V_0$ is the plasma flow velocity at the resonance point, and $k_{\text{sr}}$ is the wave vector of the seed laser, $\theta$  is the angle between the pump laser and the X-direction.

One can observe that the  reflectivity of SBS obtained by \scalebox{0.8}{\textsc{WEBS}} code agrees well with Rosenbluth gain in Fig.~\ref{inhom_fig} (c). To demonstrate that the WEBS code is applicable for simulating SBS in inhomogeneous plasmas, in Fig. \ref{inhom_fig} (d), the SBS gains at different laser intensities obtained from WEBS code simulations are represented by red diamonds. They show excellent agreement with the Rosenbluth theoretical gain. We should emphasize that the SBS gain in the WEBS simulation shown in Fig. \ref{inhom_fig} (d) originates from the central position of the Gaussian beam.

\section{Steady-State Analysis of the Interaction between CBET and SBS}\label{SBS cbet}

\begin{figure}[htbp]
    \begin{center}
      \includegraphics[width=0.52\textwidth,clip,angle=0]{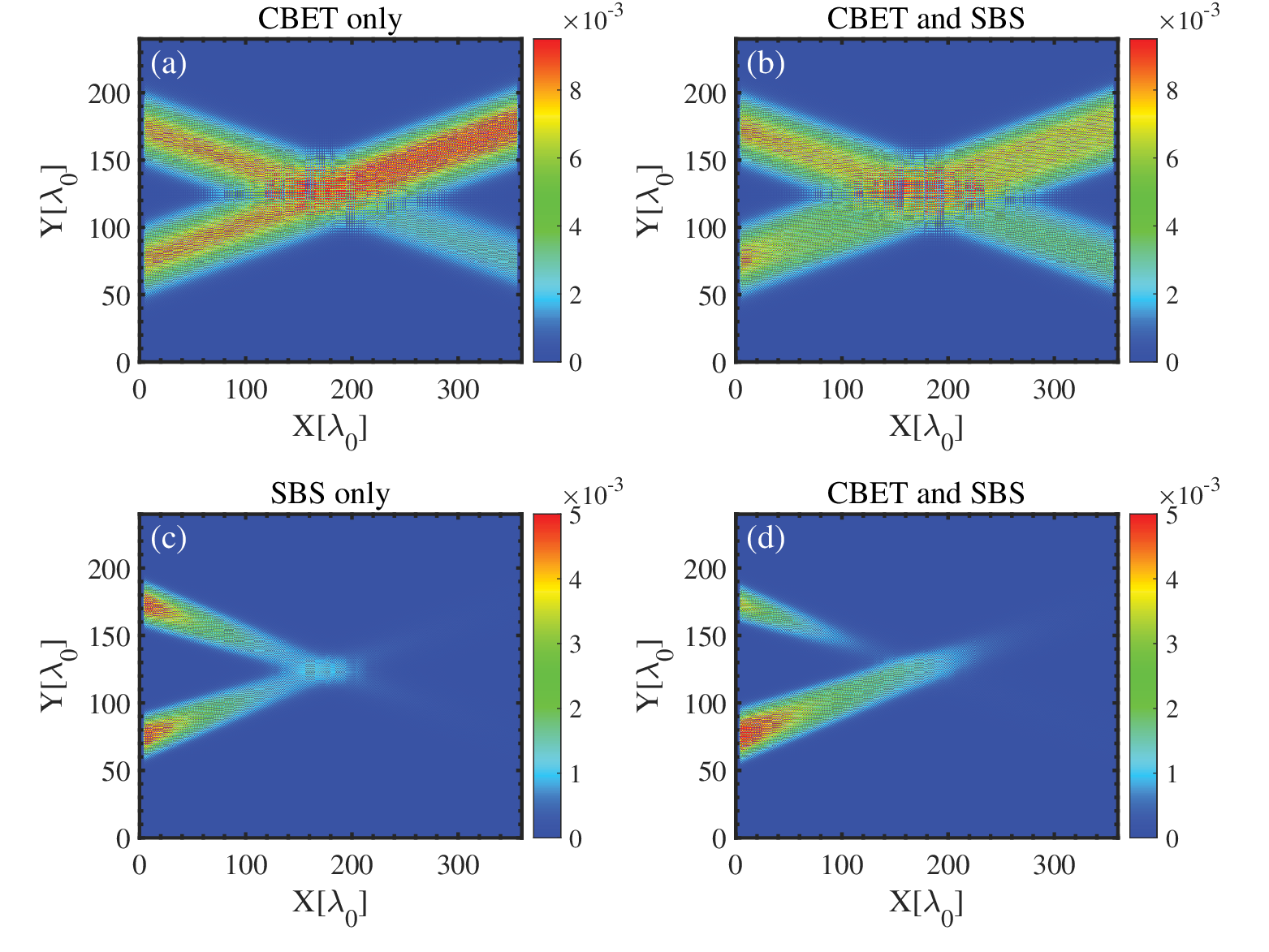}\vspace{-10pt}
      % Here is how to import EPS art [width=20mm,height=10mm][width=9cm,trim=0.5cm 21cm 0 2cm]
      \caption{\label{webs_fig1} The results by WEBS simulations. (a) The amplitude of lasers at $t = 13.2 \rm{ps}$  when we only consider CBET in simulations. (b) The amplitude of lasers at $t = 13.2 \rm{ps}$  when we both consider CBET and SBS in simulations. (c)The amplitude of SBS reflected light at $t = 13.2 \rm{ps}$ when we only consider SBS in simulations.(d)The amplitude of SBS reflected light at $t = 13.2 \rm{ps}$  when we both consider CBET and SBS in simulations.}
    \end{center}
  \end{figure}

\begin{figure}[htbp]
    \begin{center}
      \includegraphics[width=0.52\textwidth,clip,angle=0]{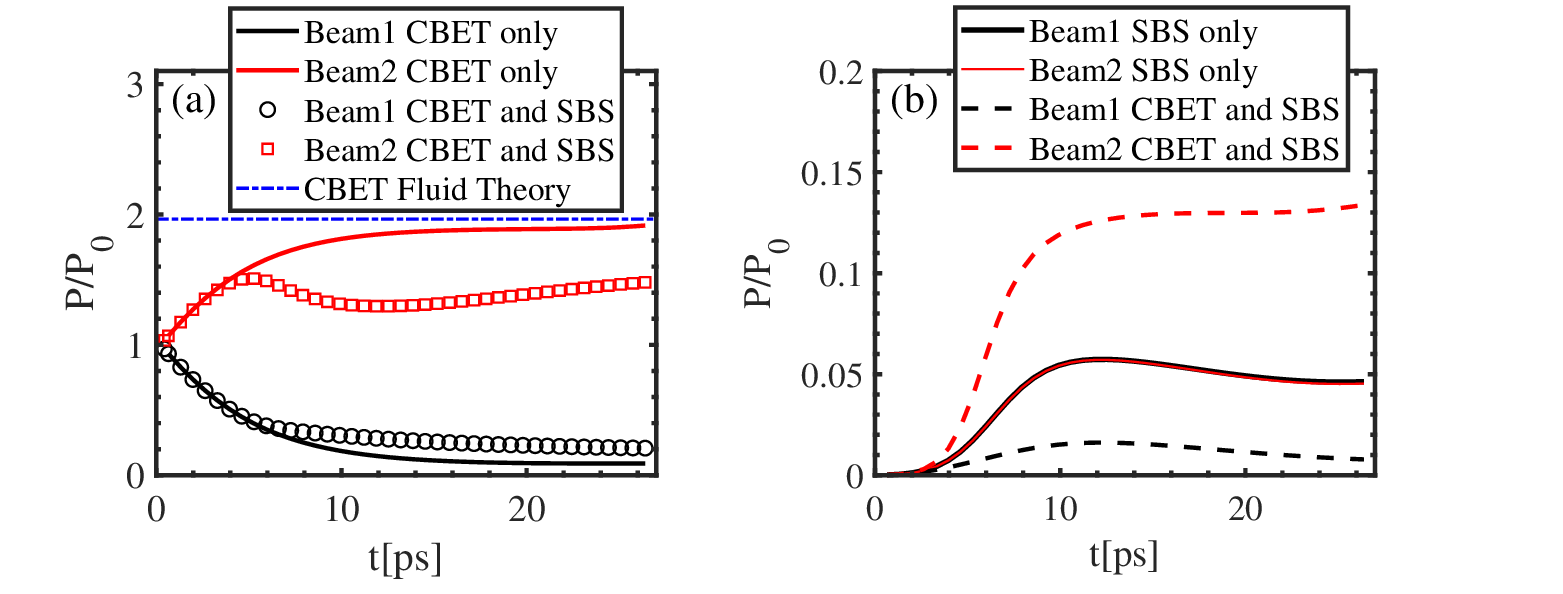}\vspace{-10pt}
      % Here is how to import EPS art [width=20mm,height=10mm][width=9cm,trim=0.5cm 21cm 0 2cm]
      \caption{\label{webs_fig3} The results of  WEBS simulations. The intensities of laser is $1\times10^{15} \rm{ W/cm^{2}}$, the electron temperature is $1\rm{keV}$,and the temperature of ions is $0.2\rm{keV}$. (a) The time-dependent lasers' energy for different cases. The red line and black line belong to the CBET only case, the blue dashed line obtained by CBET fluid theory, the red dashed line and black dashed  line belong to the CBET and SBS case. (b) The time-dependent SBS reflectivity for lasers in different cases. The red line and black line belong to the SBS only case, and the red dashed line and black dashed  line belong to the CBET and SBS case. }
    \end{center}
  \end{figure}
 The verification above demonstrates that the \scalebox{0.8}{\textsc{WEBS}} code can be employed to simulate both CBET and SBS. It is reasonable to infer that, at sufficiently high laser intensities, the interaction between CBET and SBS can cause the energy transfer associated with CBET to deviate from theoretical predictions. In this section, using the WEBS code, we demonstrate that after prolonged interaction, CBET and SBS can reach a steady state. Furthermore, we identify the parameter range in which the interaction between CBET and SBS is significant.

For simplicity, we consider two crossed laser beams with an intersection angle $\theta$ in a homogeneous, non-flowing plasma. Their frequencies satisfy the phase-matching condition for CBET, $\omega_0 = \omega_1 + \omega_2$, where $\omega_0$ and $\omega_1$ denote the frequencies of Beam~1 and Beam~2, respectively, and $\omega_2 = k_2 C_s$ represents the frequency of the ion-acoustic wave (IAW) in the CBET process. The CBET instability grows in the region where the two beams intersect. Additionally, the SBS process must be considered for each individual beam. We denote the frequency of the SBS-reflected light from Beam~1 as $\omega_3$, and the corresponding IAW frequency as $\omega_5$. For Beam~2, the respective frequencies are $\omega_4$ for the SBS-reflected light and $\omega_6$ for the IAW.

 We employ the \scalebox{0.8}{\textsc{WEBS}} code to investigate the interaction between CBET and SBS. The intensities of the two laser beams are varied in the range of \(1 \times 10^{14} \, \mathrm{W/cm}^{2}\) to \(1.3 \times 10^{15} \, \mathrm{W/cm}^{2}\), and the electron temperatures are varied from \(0.5 \, \mathrm{keV}\) to \(5 \, \mathrm{keV}\). To reduce computational time, the SBS seed level is set to \(10^{-4} I_{0}\). The total simulation time is \(26.4 \, \mathrm{ps}\), which ensures that the system reaches a steady state.

A significant advantage about WEBS is that one can easily  turn off one physical process to study the the individual evolution process of another physical process, then one turn on the physical process to study the interaction.

In Fig.~\ref{webs_fig1}, we consider laser intensities of $1 \times 10^{15}~\mathrm{W/cm^{2}}$, an electron temperature of $1~\mathrm{keV}$, and an ion temperature of $T_{e}/5$. The CBET and SBS processes interact with each other; when the system reaches a steady state, the spatial gain of CBET no longer agrees with the prediction of CBET fluid theory. For instance, in Fig.~\ref{webs_fig1}(a), where the SBS process is turned off and only CBET remains, one can clearly observe that energy is transferred from Beam~1 to Beam~2 via CBET. Similarly, as shown in Fig.~\ref{webs_fig1}(c), when only the SBS process is retained, the SBS-reflected lights for Beam~1 and Beam~2 are symmetric and exhibit equal reflectivity. However, in Fig.~\ref{webs_fig1}(b) and (d), where both SBS and CBET are active, it is evident that the presence of CBET introduces asymmetry in the SBS-reflected light between the two beams. Furthermore, the SBS process also influences the evolution of CBET.

\begin{figure}[htbp]
    \begin{center}
      \includegraphics[width=0.51\textwidth,clip,angle=0]{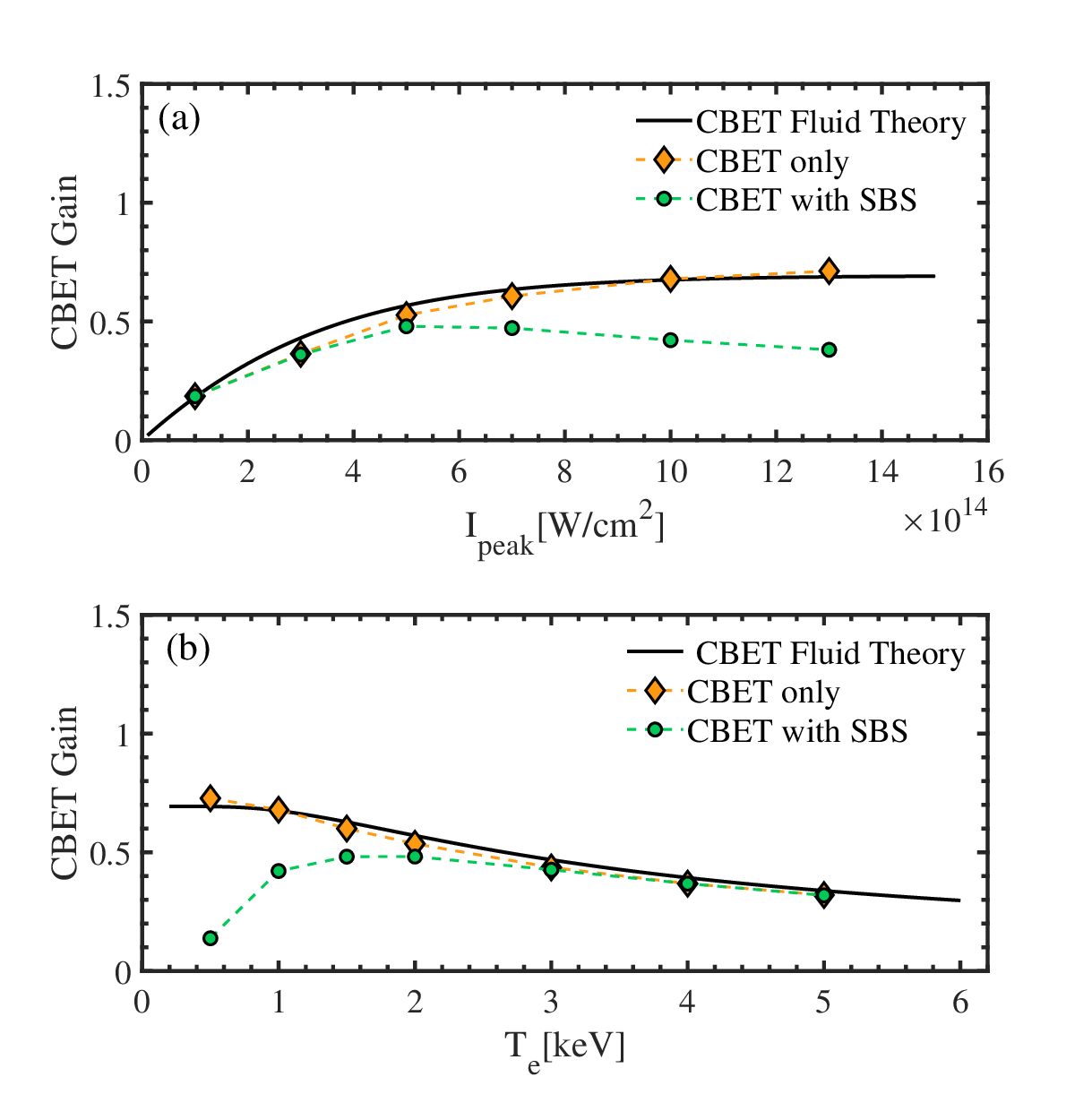}\vspace{-10pt}
      % Here is how to import EPS art [width=20mm,height=10mm][width=9cm,trim=0.5cm 21cm 0 2cm]
      \caption{\label{webs_fig4} The simulation results by WEBS. (a) The laser intensities-dependent CBET gain, the orange diamonds are the results when only consider CBET in simulations. the green dots stands for the CBET gain when both consider  CBET and SBS in simulations. The black line is obtained by CBET fluid theory. The temperatures of electrons in (a) are $1\rm{keV}$, and the temperature of ions are $0.2\rm{keV}$. (b) The plasma temperature-dependent CBET gain, the orange diamonds are the CBET gain when only consider CBET in simulations. the green dots stands for the CBET gain when both consider CBET and SBS in simulations. The black line is obtained by CBET fluid theory. The peak intensities of laser beams in (b) are  $1\times10^{15} \rm{ W/cm^{2}}$. }
    \end{center}
  \end{figure}

  Fig.~\ref{webs_fig3}(a) shows the time evolution of the laser beams' energy. The red and black curves correspond to the case with only CBET. At around $t = 20~\mathrm{ps}$, the system reaches a steady state, where the energy of Beam~2 agrees with the CBET fluid theory (blue dashed line) and the energy of Beam~1 approaches zero, which is a clear signature of pump depletion. The red squares and black circles represent the case with both CBET and SBS. In this scenario, pump depletion of Beam~1 persists, but the energy gain of Beam~2 is reduced. This reduction occurs because SBS consumes a significant portion of the energy from Beam~2.

  Fig~\ref{webs_fig3}(b) shows the time evolution of the SBS reflectivity for each laser beam. The solid red and black lines correspond to the case with only SBS, in which the reflectivities of Beam~1 and Beam~2 are identical, consistent with theoretical expectations. The dashed red and black lines represent the case with both CBET and SBS active. In Fig.~\ref{webs_fig3}(b), if the interaction between CBET and SBS is neglected, the average total SBS reflectivity for both Beam~1 and Beam~2 is approximately $4\%$. When the CBET-SBS interaction is included, the time-averaged SBS reflectivity for Beam~1 is $1.07\%$, while for Beam~2 it is $10.27\%$. The increase in total SBS reflectivity is attributed to the interplay between the CBET and SBS processes.

  Next, we examine the spatial gain of CBET under varying laser and plasma parameters. As shown in Fig.~\ref{webs_fig4}(a), the orange diamonds correspond to the simulation case where only CBET is considered. The CBET gain increases with laser intensity; when the peak laser intensity reaches $1.3 \times 10^{15} \, \mathrm{W/cm^{2}}$, the intensity of Beam~2 is nearly doubled due to pump depletion. These simulation results for the CBET-only case agree well with the prediction of CBET fluid theory, which is represented by the black line.

  In contrast, when the interaction between CBET and SBS is included (green dots in Fig.~\ref{webs_fig4}(a)), the CBET gain begins to deviate from the fluid-theory prediction once the laser intensity exceeds $6 \times 10^{14} \, \mathrm{W/cm^{2}}$. This behavior is consistent with our earlier analysis, as CBET enters the depletion regime and the concurrent growth of SBS depletes part of the energy in Beam~2.

  We also perform fluid simulations under different plasma temperatures, as illustrated in Fig.~\ref{webs_fig4}(b). We note that for the results shown in Fig.~\ref{webs_fig4}(b), the ion temperature is set as $T_i = 0.2\,T_e$, and the laser intensities are fixed at $1 \times 10^{15}~\mathrm{W/cm^{2}}$. The orange diamonds represent the results from simulations that consider only CBET, which show good agreement with the predictions of CBET fluid theory. When the plasma temperature is below $2~\mathrm{keV}$, the CBET gain deviates from the fluid theory. In this regime, Beam~2 gains substantial energy from CBET, causing the SBS process to grow more rapidly in the region downstream of the CBET interaction. The interplay between CBET and SBS thus leads to the observed discrepancy between the simulation results and the fluid theory. As the temperature increases, CBET transitions from the depletion regime into a linear or weakly interacting regime, allowing CBET and SBS to develop independently. This explains why the CBET regains agreement with the fluid theory at higher temperatures.

\section{Conclusion and discussion}\label{conclusion}

In this work, we  develop a wave-based simulation model, named the \scalebox{0.8}{\textsc{WEBS}} code, to study laser-plasma instabilities. The program employs the Du Fort-Frankel scheme to solve the nonlinear Schr\"odinger equations, achieving good numerical accuracy even with a coarse spatial mesh. The code has been validated against PIC simulations in the fluid regime. We then investigate the interaction between CBET and SBS in a homogeneous plasma. It is found that in the low-intensities regime, SBS growth does not significantly affect the spatial gain of CBET. However, in the depletion regime, the CBET gain is notably influenced by SBS. This interplay leads to a pronounced asymmetry in SBS reflectivity between Beam~1 and Beam~2, and the high reflectivity in turn suppresses the CBET gain. These theoretical predictions are consistently confirmed by our \scalebox{0.8}{\textsc{WEBS}} simulations. Overall, our study provides a practical framework for identifying the parameter regimes in which CBET and SBS interact strongly, offering guidance for the interpretation of experiments and the design of inertial confinement fusion scenarios.

In inertial confinement fusion experiments, hundreds of laser beams simultaneously irradiate the hohlraum. In the overlapping regions of these beams, the absolute instability threshold of SBS is lowered. Particularly in inhomogeneous plasmas, the side-scattering of SBS may transition into an absolute instability mode. Our WEBS code is capable of simulating SBS side-scattering. In future work, we will employ the \scalebox{0.8}{\textsc{WEBS}} code to investigate the characteristics of SBS side-scattering in inhomogeneous plasmas.

\section{Acknowledgements}
We are pleased to acknowledge useful discussions with  Y. G. Chen, S. Tan and Y. X. Li.  This work was supported by the Scientific Research Foundation for High-level Talents of Anhui University of Science and Technology (Grant No.2022yjrc106), National Natural Science Foundation of China (Grant Nos.12405277 and 12475237), Anhui Provincial Natural Science Foundation (Grant No.2308085QA25), the Strategic Priority Research Program of Chinese Academy of Sciences (Grant No.XDA25050700 and XDA25010100) and the Fund of National Key Laboratory of Plasma Physics (Grant No.6142A04230103).

%\bibliographystyle{apsrev4-1}
%\bibliography{ClassicSpinModels}

\begin{thebibliography}{100}
\newcommand{\DOI}[1]{doi: \href{https://doi.org/#1}{#1}}

\bibitem{ICF1}R. Betti, et al.,  Phys. Rev. Lett. {\bf 98}, 155001 (2007).(\DOI{10.1103/PhysRevLett.98.155001})


\bibitem{ICF3}Max Tabak, et al.,  Phys. Plasmas {\bf 1}, 1626 (1994).(\DOI{10.1063/1.870664})
\bibitem{D_D}  S. E. Bodner, et al., Phys. Plasmas {\bf 5}, 1901 (1998).(\DOI{10.1063/1.872861})
\bibitem{I_D}   J. Lindl, Phys. Plasmas {\bf 2}, 3933 (1995).(\DOI{10.1063/1.871025})
\bibitem{ICFn}V. Tikhonchuk, et al., Matter Radiat. Extremes {\bf 4}, 045402 (2019).(\DOI{10.1063/1.5090965})
\bibitem{Icfmre} M. Q. Fan, et al., Matter Radiat. Extremes {\bf 8}, 025902 (2023).(\DOI{10.1063/5.0129434})
\bibitem{Icfmre2} Y. Guo, et al., Matter Radiat. Extremes {\bf 8}, 035902 (2023). (\DOI{10.1063/5.0136567})
\bibitem{Icfmre3} T. Gong, et al.,  Matter Radiat. Extremes {\bf 4}, 055202 (2019).(\DOI{10.1063/1.5092446})
 \bibitem{Icfmre4} S. Tan,  et al., Matter Radiat. Extremes {\bf 9}, 057402 (2024).(\DOI{10.1063/5.0206740})



\bibitem{cbet5} A. Oudin, et al.,  Phys. Rev. Lett.,  {\bf 127}, 265001 (2021).(\DOI{10.1103/PhysRevLett.127.265001})

\bibitem{cbet6} A. G. Seaton, et al., Phys. Plasmas, {\bf 29}, 042706 (2022).(\DOI{10.1063/5.0078800})
\bibitem{cbet7} A. G. Seaton,et al., Phys. Plasmas, {\bf 29}, 042707 (2022).(\DOI{10.1063/5.0078801})
\bibitem{cbet8} A. M. Hansen, et al., Phys. Rev. Lett.,  {\bf 126} 075002 (2021).(\DOI{10.1103/PhysRevLett.126.075002})
\bibitem{cbet9} K. L. Nguyen, et al., Phys. Plasmas, {\bf 28}, 082705 (2021). (\DOI{10.1063/5.0054008})




\bibitem{sbs_m_l} Q. Wang, et al., Matter Radiat. Extremes {\bf 8}, 055602 (2023).(\DOI{10.1063/5.0151372})

\bibitem{sbs1} P. Neumayer, et al., Phys. Rev. Lett. {\bf 100}, 105001 (2008).(\DOI{10.1103/PhysRevLett.100.105001})
\bibitem{sbs2} J. Li, et al., Phys. Rev. E {\bf 101}, 033206 (2020).(\DOI{10.1103/PhysRevE.101.033206})
\bibitem{sbs3} S. Zhang, et al., Phys. Rev. E {\bf 103}, 063208 (2021).(\DOI{10.1103/PhysRevE.103.063208})
\bibitem{sbs4} H. A. Rose and D. F. DuBois, Phys. Rev. Lett. {\bf 72}, 2883 (1994).(\DOI{10.1103/PhysRevLett.72.2883})
\bibitem{sbs5} C. Z. Xiao, et al., Phys. Rev. E {\bf 104}, 065203 (2021).(\DOI{10.1103/PhysRevE.104.065203})
\bibitem{HaoL1} L. Hao, et al., Phys. Plasmas {\bf 24}, 062709 (2017). (\DOI{10.1063/1.4989702})

\bibitem{press1} J. Nuckolls, et al., Nature (London) {\bf 239}, 139 (1972).(\DOI{10.1038/239139a0})
 \bibitem{press2} J. D. Lindl, et al., Phys. Plasmas {\bf 11}, 339 (2004). (\DOI{10.1063/1.1578638})


\bibitem{IV}I. V. Igumenshchev, et al., Phys. Plasmas, {\bf 17}, 122708 (2010).  (\DOI{10.1063/1.3532817})
\bibitem{cbet1} E. I. Moses, et al., Phys. Plasmas, {\bf 16}, 041006 (2009).(\DOI{10.1063/1.3116505})

\bibitem{cbet2}P. Michel, et al., Phys. Rev. Lett., {\bf 102}, 025004 (2009).(\DOI{10.1103/PhysRevLett.102.025004})
\bibitem{cbet3}P. Michel, et al., Phys. Plasmas,  {\bf 17}, 056305 (2010). (\DOI{10.1063/1.3325733})
\bibitem{cbet10} D. J. Stark, et al., Phys. Plasmas,  {\bf 28}, 022702 (2021).(\DOI{10.1063/5.0022091})
\bibitem{cbet11} D. J. Stark, et al., Phys. Plasmas,  {\bf 30}, 042714 (2023).(\DOI{10.1063/5.0134881})
\bibitem{cbet12}L. Yin, et al., Phys. Plasmas,  {\bf 30}, 042706 (2023).(\DOI{10.1063/5.0134867})



\bibitem{epoch} T. D. Arber, et al., Plasma Phys. Control. Fusion {\bf 57}, 113001 (2015). (\DOI{10.1088/0741-3335/57/11/113001})

\bibitem{pic2} X. H. Yang, et al., Matter Radiat. Extremes {\bf 8}, 035901 (2023).(\DOI{10.1063/5.0137973})
 \bibitem{jiao} J. L. Jiao, et al.,  Plasma Sci. Technol. {\bf 24},  105201(2022).(\DOI{10.1088/2058-6272/ac74a8})



\bibitem{lpse} J. F. Myatt, et al., Journal of Computational Physics {\bf 399},  108916 (2019). (\DOI{10.1016/j.jcp.2019.108916})
\bibitem{cbet13} J. W. Bates, et al. Phys. Rev. E {\bf 97}, 061202(R) (2018).(\DOI{10.1103/PhysRevE.97.061202})
\bibitem{ray1}R. K. Follett, et al.,Phys. Plasmas {\bf 29}, 113902 (2022)(\DOI{10.1063/5.0123462})
\bibitem{ray2}R. K. Follett,et al.,Phys. Plasmas 30, 042102 (2023)(\DOI{10.1063/5.0137420})
\bibitem{jiaxb} X. Jia, Q. Jia, J. Xiao, and J. Zheng, Phys. Plasmas {\bf 32}, 062708 (2025). (\DOI{10.1063/5.0273367})
 \bibitem{har}  S. H$\ddot{u}$ller, et al. Phys. Plasmas {\bf 13}, 022703 (2006). (\DOI{10.1063/1.2168403})
\bibitem{raj} G. Raj, et al., Phys. Rev. Lett., {\bf 118}, 055002 (2017).(\DOI{10.1103/PhysRevLett.118.055002})











\bibitem{wangqing2} Q. Wang, et al., Plasma Phys. Control. Fusion {\bf 61}, 085017 (2019).(\DOI{10.1088/1361-6587/ab2736})

\bibitem{wangqing1}  Q. Wang, et al., Plasma Phys. Control. Fusion {\bf 60}, 025016 (2018).(\DOI{10.1088/1361-6587/aa98bb})

\bibitem{rosenbluth1} M. N. Rosenbluth, Phys. Rev. Lett. {\bf 29}, 565 (1972). (\DOI{10.1103/PhysRevLett.29.565})
\bibitem{rosenbluth2} M. N. Rosenbluth, et al., Phys. Rev. Lett. {\bf 31}, 1190 (1973).(\DOI{10.1103/PhysRevLett.31.1190})














\bibitem{shift1} Williams E A, et al.,  Phys. Plasmas {\bf 11} 231(2004)(\DOI{10.1063/1.1630573})

\bibitem{shift}  R. L. Berger, et al.,  Phys. Plasmas {\bf 20}032107 (2013).(\DOI{10.1063/1.4794346})

\bibitem{TPD}  private communication with Y. G. Chen.

\bibitem{DFF} E. C. Du Fort and S. P. Franke, Mathematical Tables and Other Aids to Computation,{\bf 7} p135 (1953). (\DOI{10.2307/2002754})

\bibitem{DFF2} C. M. Campbell and P. Keast., Mathematics of Computation, {\bf 22}, pp. 336$-$346 (1968).
(\DOI{10.2307/2004663})

\bibitem{DFF3} X. F. Yang, D. A. Ralescub, Mathematics and Computers in Simulation, {\bf 181}, pp. 98$-$112 (2021).
(\DOI{10.1016/j.matcom.2020.09.022})


\bibitem{DFF5}  X. D. Hang, Mathematica Numerica Sinica, {\bf 37}: 273$-$285 (2015).
(\DOI{10.12286/jssx.2015.3.273})

\bibitem{DFF4} L. X. Wu, SIAM Journal on Numerical Analysis, {\bf 33}, pp. 1526$-$1533 (1996).
(\DOI{10.1137/S0036142994270636})



\bibitem{xin1} R.D. Ruth, IEEE Trans. Nuclear Science, {\bf 30}, 2669$-$2671 (1983).
(\DOI{10.1109/TNS.1983.4332919})

\bibitem{xin2}D. Donnelly, E. Rogers, Am. J. Phys., {\bf 73}, 938$-$945 (2005).
\bibitem{tang} C. L. Tang, J. Appl. Phys., {\bf 37}, 2945 (1966).(\DOI{10.1063/1.1703144})


\end{thebibliography}

\end{document}